# Scaling of broadband Ho:CALGO regenerative amplifier to multi-mJ pulse energy

ANNA SUZUKI,[1,*] MICHAEL MÜLLER,[1] SERGEI TOMILOV,[1] CLARA. J. SARACENO[1,2]

[1]*Photonics and Ultrafast Laser Science, Ruhr-Universität Bochum, Universitätsstrasse 150, 44801 Bochum,Germany*
[2]*Research Center Chemical Science and Sustainability, University Alliance Ruhr, 44801 Bochum, Germany*
*\*anna.ono@ruhr-uni-bochum.de*

**Abstract:** We report on energy scaling of a 2.08-μm wavelength regenerative amplifier (RA) system based on the broadband gain material Ho:CaAlGdO$_4$ (CALGO) to multi-mJ pulse energy at kHz repetition rates. Compared to previous reports, energy scaling was enabled thanks to an upgraded seed laser with a higher fluence and better spectral overlap to the gain spectrum of Ho:CALGO, which increased amplification efficiency. Bifurcation-free energy extraction was investigated experimentally and numerically for various repetition rates. A stable output was obtained at 10 W average power for repetition rates of 30 kHz and above. In addition, stable 3.4-mJ energy extraction was achieved at a 1-kHz repetition rate. We discuss the further scaling potential of pulse energy and pulse duration.

## 1. Introduction

In recent years, high-power and high-energy ultrafast laser sources emitting in the vicinity of 2 μm wavelength are in growing demand due to several advantages in various applications. For example, processing applications of narrow-bandgap semiconductor materials such as silicon or germanium benefit from longer wavelengths for backside modification or in-volume processing [1,2]. Furthermore, many conversion processes benefit from longer driving wavelength lasers: for instance, high-harmonic generation requires longer wavelength drivers to reach the soft X-ray region in the water window [3,4], and terahertz (THz) sources based on nonlinear crystals can withstand higher intensities due to smaller driving photon energies [5].

Today, many 2 μm wavelength sources rely on optical parametric conversion of high-energy Ti:sapphire lasers or high-average power Yb-based lasers [6,7]. However, direct laser emission in this wavelength range is a simpler alternative for overall efficiency and repetition rate scaling. At higher pulse energies, ultrafast 2-μm lasers using thulium (Tm) and holmium (Ho)-doped active media have been a topic of active research and have seen significant progress in recent years. Tm-doped fiber amplifiers are excellent for average power scaling due to the cooling efficiency of the fiber geometry. For example, an ultrafast Tm fiber amplifier was demonstrated to deliver 1-kW average power with 265-fs pulse duration at 80-MHz repetition rate [8]. However, their typical operation wavelength of 1.95 μm suffers from water vapor absorption, imposing challenges for free-space beam transportation and beam quality. In addition, high nonlinearity and critical peak power for self-focusing limit pulse energy scaling. Still, 1.65-mJ energy amplification was achieved at 101 kHz, however, coherent beam combining was required to achieve high peak power, which made the system rather complex.

In this regard, Ho-based bulk laser amplifiers are a promising alternative for power and energy scaling in this spectral region. Their large cross sections and millisecond-level upper-level lifetime ensure a large energy storage capability, and the small quantum defect under in-band pumping results in a small heat load. Moreover, their central emission wavelength of 2.1 μm is in an atmospheric transmission window, which is beneficial for beam transport and power scaling. For example, a high average power of 111 W was recently achieved by a Ho:YAG RA followed by two booster stages, delivering 334 ps pulses at 10 kHz in a non-CPA configuration [9]. For the CPA arrangement, Ho:YLF currently serves as the main workhorse for high-power and high-energy systems due to a slightly broader bandwidth compared to Ho:YAG, which enables few-ps pulse durations. In addition, its small nonlinear refractive index and its negative and small thermo-optic coefficient are desirable for high-power and high-energy operation [10,11]. A pulse energy of 75 mJ with a pulse duration of 2.2 ps was reported in Ho:YLF RA followed by two booster amplifier stages at 1 kHz [12]. A RA and a cryogenically cooled booster amplifier were also demonstrated, achieving 260 mJ of pulse energy with a pulse duration of 16 ps at a lower repetition rate of 100 Hz [13].

However, sub-ps pulse amplification is still challenging with conventional Ho-doped materials due to their narrow and structured gain profile. In this regard, disordered host materials for the gain medium are a promising way to realize broadband amplification while maintaining the advantages of Ho laser emission properties. In particular, the broadband material Ho:CALGO has recently demonstrated to be a very promising material for broadband Ho laser systems. Its disordered structure causes inhomogeneous spectral broadening, resulting in a broad and flat gain profile, thus allowing for both record-short pulse and high-power operation from oscillators [14,15] but also mitigating gain narrowing in high-gain amplifiers [16–18]. Moreover, CALGO exhibits exceptionally high thermal conductivity among disordered host materials [19], which makes it suitable for high-average power laser operation. Previously, we reported the first demonstration of a Ho:CALGO RA seeded by a home-built 2093-nm Tm,Ho:CLNGG laser, which delivers 110-µJ pulses with a pulse duration of 750 fs at 100 kHz [20].

In this study, we improved the seed source by increasing its power and adjusting the center wavelength for the gain peak of Ho:CALGO, resulting in more efficient energy extraction. In addition, we demonstrate energy scaling of the amplifier system to the mJ level. In continuous wave (CW) pumped RA systems, bifurcation (multifurcation) instability occurs when the repetition rate is greater than the inverse of the upper-level lifetime of the gain medium [21]. Ho:CALGO has a fluorescence lifetime of 5.28 ms [22] corresponding to 189 Hz, therefore, operation of the amplifier in the kilohertz repetition rate region needs to be carefully designed to avoid these unstable regimes of operation. Using a numerical simulation, we predicted the onset of bifurcation instability at different repetition rates and experimentally confirmed stable amplifier operation at higher repetition rates of 30, 40, 50 kHz with a 10-W average output power and sub-ps pulse duration. Moreover, 3-mJ pulse energy was obtained with a pulse duration of 1.23 ps at a lower repetition rate of 1 kHz. This work demonstrates the first mJ-class amplifier system using a disordered gain medium in the 2-µm spectral range.

## 2. Experimental setup of the Ho:CALGO regenerative amplifier

### 2.1 Seed laser oscillator

An important consideration in the design of efficient high-gain amplifiers is that the seed spectrum matches as closely as possible the gain spectrum of the amplifier medium for efficient energy extraction. Since Ho:CALGO lasers operate in a quasi-three-level laser system, the peak gain wavelength changes significantly with the inversion level. This is clearly seen in the fact that while mode-locked Ho:CALGO oscillators operating at low inversion level emit at around 2120 nm [14,15], the gain of Ho:CALGO amplifiers at higher inversion is maximized around 2077 nm [22], making Ho:CALGO seeders not the ideal choice. To circumvent this problem, we previously utilized a Tm,Ho-codoped CLNGG laser as a seed source operating at a center wavelength of 2093 nm [20]. However, this laser system was very limited in output power and pulse energy of 91 mW and 1.3 nJ, respectively, due to poor conversion efficiency of Tm,Ho codoped laser system [23,24], and thus still represented a limitation for high-energy operation of the amplifier. In this new design, Tm:Lu$_2$O$_3$ is used for the seed laser gain medium, with a center wavelength in mode-locked operation typically around 2070 nm and with hundreds of mW power levels [25,26]. The experimental setup of the seed oscillator is depicted in Fig. 1(a). The gain medium is a 4-at.% doped Tm:$Lu_2O_3$ ceramic with a length of 6 mm, pumped by a 793-nm multi-mode fiber-coupled diode with a core diameter and a numerical aperture of 106.5 μm and 0.15, respectively. A semiconductor saturable absorber mirror (SESAM) (RefleKron Ltd.) is used to start and stabilize mode-locking with a saturation fluence of 10 μJ/$cm^2$, a modulation depth of 0.23%, and a nonsaturable loss of 0.12% measured at 2130 nm [14]. The total cavity group delay dispersion (GDD) was set to -2200 $fs^2$. The mode-locked laser delivers soliton pulses with an average power of 230 mW and a pulse energy of 3.2 nJ at a repetition rate of 71.6 MHz. The center wavelength is 2085 nm with a full width at half maximum (FWHM) bandwidth of 15.1 nm as shown in Fig. 1(b). The pulse duration was 300 fs, which is close to the Fourier transform limit assuming $sech^2$ pulses [Fig. 1(c)]. As a result, we improved the pulse energy of the seed laser by a factor of 2.5 and enabled better overlap of seed and amplifier operating wavelengths, as shown in Fig. 1(b).

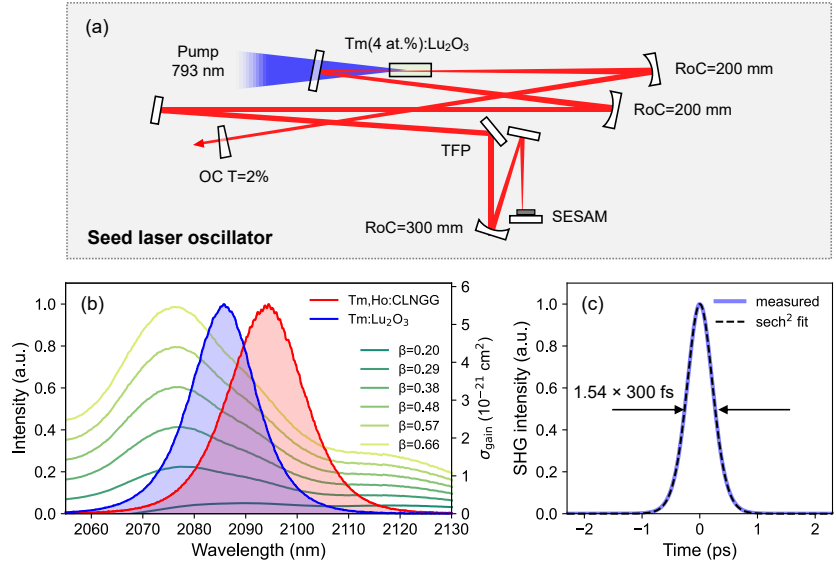

Fig. 1. (a) Experimental setup of the seed laser oscillator (TFP: thin-film polarizer, OC: output coupler, RoC: radius of curvature). (b) Optical spectra of the mode-locked Tm,Ho:CLNGG laser (red) and Tm:$Lu_2O_3$ laser (blue), and gain spectra of Ho:CALGO for π-polarization (green). (c) Autocorrelation trace of the Tm:$Lu_2O_3$ mode-locked laser.

## 2.2 Testing the influence of the seed in CPA setup

We first compared the amplifier performance with the previously used 2093-nm seed laser and the new 2085-nm seed laser. For a fair comparison, only the seed laser was changed, and otherwise we used the same 220-ps stretched pulse duration and the same 27-mm long Ho:CALGO crystal as in our previous work [20]. The setup is similar to the one shown in Fig. 3, with small differences in the line constant of gratings and crystal length. For this test, we performed pulse amplification at a 100-kHz repetition rate, where we already demonstrated the first results. Figure 2(a) shows the energy evolution of the amplifier, and pulse energy of 100 μJ (10-W average power) was obtained with both seed lasers. With the 2085-nm seed, an RT number of 21 was needed to extract the energy instead of 28 in the 2093-nm seed case, indicating improved amplification efficiency. However, the spectrum became narrower than with the 2093-nm seed [Fig. 2(b)]. In the 2093-nm seed case, the shorter wavelength side experienced more amplification due to a larger gain, and the longer wavelength side still was moderately amplified, i.e., amplification for the whole spectral range was eventually balanced with this gain factor, resulting in a broader spectral bandwidth. The output in the case of the 2085-nm seed features slightly less bandwidth but sub-ps pulse duration was obtained after compression, as shown in Fig.2(c). Since the higher seed fluence allows to increase the bifurcation threshold at lower repetition rates [25,26], as well as to decrease B-integral by reducing RT number, we employ the 2085-nm seed laser for the following experiments aiming to maximize output pulse energy.

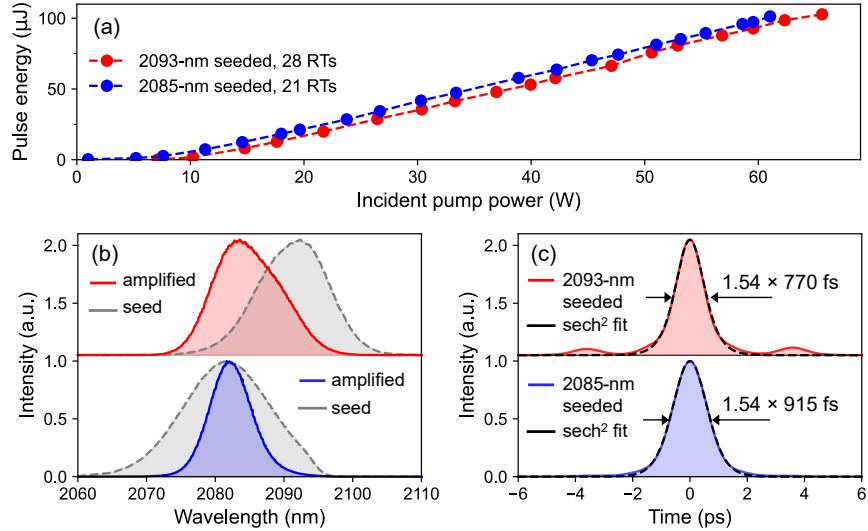

Fig. 2 Comparison of the Ho:CALGO RA performance with different wavelength seed lasers. (a) Pulse energy as a function of incident pump power at 100 kHz. (b) Optical spectra of the seed after the stretcher (dashed), amplified pulses (solid). (c) Autocorrelation traces of the amplified pulses after the compression.

## *2.3 CPA setup with upgraded seed for energy scaling*

The experimental setup of the CPA system is shown in Fig. 3. The seed pulses generated from the mode-locked Tm:Lu$_2$O$_3$ laser are sent to a stretcher via an isolator to prevent feedback. The pulses are stretched by a grating pair with a line constant of 800 line/mm in the Treacy configuration, which gives GDD of -72 ps$^2$, resulting in a 500-ps pulse duration. The gain medium of the RA stage is a Brewster's cut 24.5-mm Ho(1 at.%):CALGO crystal with π-polarization geometry (E||c). It is pumped by a 1908-nm single-mode continuous wave (CW) Tm-doped fiber laser. After the amplification, pulses are compressed using a Martinez compressor with the same grating constant as the stretcher and concave mirrors with a radius of curvature of 1500 mm.

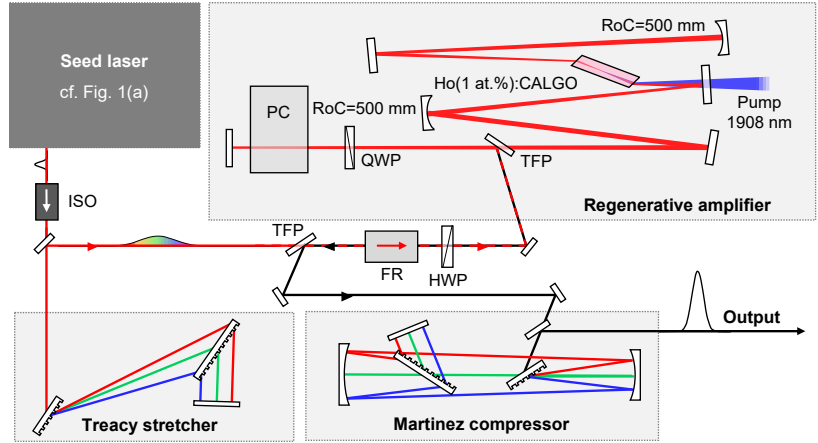

Fig. 3. Experimental setup of the Ho:CALGO CPA, including a seed laser oscillator, a stretcher, a regenerative amplifier, and a compressor (ISO: isolator, TFP: thin-film polarizer, FR: Faraday rotator, Q/HWP: quarter-/half-wave plate, PC: Pockels cell).

## 3. Numerical simulation and experiments for energy scaling

### 3.1 Numerical simulations for the prediction of bifurcation instability

Bifurcation instability is the bottleneck for pulse-to-pulse stability in CW-pumped RA systems that operate at higher repetition rates than the inverse upper-level lifetime. The energy bifurcation occurs owing to the imbalance between the gain depletion by the seed and the gain recovery by the pump [27]. Therefore, the dynamics of energy evolution in the RA are defined by the balance of all relevant parameters, such as the repetition rate, number of roundtrips, pump intensity, and seed fluence. Kroetz *et al.* demonstrated a comprehensive numerical study revealing the existence of 4 different operation regimes in the RA based on the relationship between the peak of the average pulse energy [yellow stars in Fig. 4(a)] and the onset of bifurcation [28]. Regime 1): bifurcation-free regime. Regime 2): bifurcation will be seen at early RTs, however, the stable operation regime at high energy levels appears beyond the bifurcation instability. Regime 3): bifurcation will be seen at most RTs, stable operation is only available with a low energy level before kicking off bifurcation. 4) Maximum pulse energy can be extracted with stable operation, and bifurcation starts at a later point. For efficient energy extraction, regimes 1) and 4) are ideal, and regime 2) is still possible to extract high pulse energy at the expense of increased B-integral.

We carried out a numerical simulation to predict the operation regimes of our amplifier system. We used the simulation model presented in [29], based on a spectrally resolved Frantz-Nodvik (FN) equation. The input parameters are summarized in Table 1. We simulated the operation regimes by variation of the repetition rate and pump intensity to find an operation point allowing us to extract high energy with stable output.

Table 1. Input parameters used in the simulation

| Parameter | Value | Unit |
|---|---|---|
| Pump and laser spot size | 145×270 | $\mu m^2$ |
| Pump power | <100 | W |
| Pump wavelength | 1908 | nm |
| Seed energy | 2.37 | µJ |
| Doping concentration | $1.25 \times 10^{20}$* | $cm^{-3}$ |
| Fluorescence lifetime of $^5I_7$ manifold of Ho(1%):CALGO | 5.27 [22] | ms |

| Absorption/emission cross sections | [22] |

*Calculated with a cation density of $1.25 \times 10^{22}$ cm$^{-3}$ with respect to the Gd sites [30]

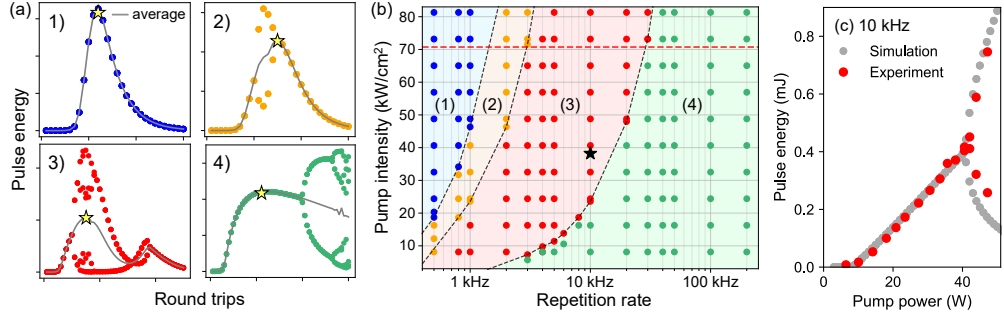

Fig. 4. (a) Examples for the typical energy evolutions in the four operation regimes as a function of RTs. (b) Numerically simulated operation regimes of the Ho:CALGO RA as a function of pump intensity and repetition rate. (c) Comparison of simulation and experiment at 10 kHz with 26 RTs.

Figure 4 (b) shows the simulated operation regimes. The red dashed line shows the maximum pump intensity, which we used in the following experiments. To verify the simulation, we carried out an experiment at 10 kHz [a black star in Fig. 4(b)]. Figure 4(c) shows simulated and measured energy evolution with 26 RTs as a function of pump power. The energy bifurcation was predicted to start around 41-W of pump power, and the experimental result of energy evolution and the bifurcation starting point matched well, indicating that our bifurcation prediction worked. Based on the bifurcation map, high energy extraction without bifurcation instability would be possible above 30 kHz. Our initial result at 100 kHz belongs to this region. Also, the bifurcation-free regime 1) appears below 1 kHz at high pump intensity levels. Therefore, we performed amplifier experiments at these repetition rates.

*3.2 Experimental results of the amplifier at kilohertz repetition rate*

We first performed amplification experiments in regime 4)[Fig. 4(a)] with the results shown in Fig. 5. At 30, 40, and 50 kHz repetition rates, we obtained an average power of 10 W after 26 RT for an incident pump power of 90 W, corresponding to pulse energies of 333, 250, and 200 µJ, respectively. Compared to our previous result [20], the result obtained at 30 kHz represents a 3.3 times improvement of pulse energy at constant average output power, further demonstrating the potential of Ho:CALGO. The power slopes started saturating above 85 W of pump power, thus, we did not increase beyond 90 W to avoid the risk of crystal damage. Here, the simulation showed that the operation point at our maximum pump power at 30 kHz lies close to the boundary between operation regimes 3) and 4) [Fig. 4(b)], however, energy bifurcation was not observed at 30 kHz within the range of pump powers used. That was confirmed by measuring the pulse amplitude fluctuation using a fast photodiode and an oscilloscope, and we measured the pulse-to-pulse fluctuation of ≈2%, which is comparable to that observed at 40 and 50 kHz. Figure 5(c) shows the corresponding optical spectra at 10-W average power, which are almost identical for all repetition rates. The center wavelength was 2085 nm with a spectral bandwidth of 7.9 nm. The compressed pulse duration was 950 fs for 50 kHz, and 970 fs for 40 and 30 kHz. In the operation regime 4), we achieved increasing pulse energies with 10-W average power by reducing the repetition rate with sub-ps pulse durations.

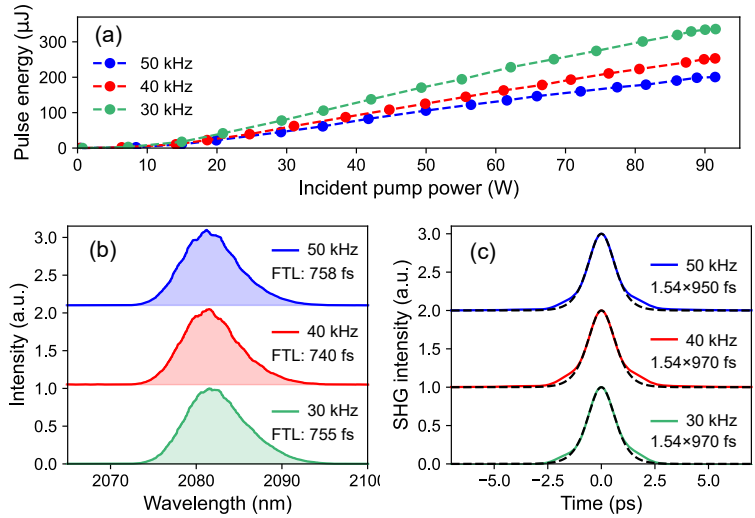

Fig. 5. (a) Pulse energies of the Ho:CALGO RA as a function of incident pump power at 50 (blue), 40 (red), and 30 kHz (green). (b) Optical spectra of amplified pulses, and (c) autocorrelation traces of compressed pulses.

Next, amplifier operation in regime 1) was investigated at 1 kHz. First, the pulse evolution was simulated to find the optimum number of RTs at 90-W pump power, which we set as a limit in the previous experiment. The simulation predicted that only 14 RTs are required to saturate the gain at this pump power as shown in gray Fig. 6(a). In this case, no bifurcation was predicted in the whole pump power range. Subsequently, we performed the amplifier experiment at 14 RTs and 1-kHz repetition rate. We obtained a maximum average power of 3.4 W at 57-W pump power, corresponding to 3.4-mJ pulse energy. Further scaling was limited by damage to the gain crystal, most likely caused by the amplified pulse itself, which remains to be investigated.

The compressed pulses were characterized using a spectrometer and a home-built second-harmonic frequency-resolved optical gating (SH-FROG), as shown in Fig.6 (b-e). The retrieved pulse duration from the FROG trace was 1.23 ps with a residual higher-order phases remaining, while the FTL pulse duration calculated from the measured spectrum is 1.05 ps. The SHG intensity autocorrelator measured the pulse duration of 1.30 ps, assuming a $sech^2$ shape. The transmission of the compressor was 75%, which is limited by the available grating diffraction efficiency, resulting in a pulse energy of 2.55 mJ, and a peak power was calculated to be 1.77 GW. The compression efficiency can be increased by switching the grating to a higher transmission one, which was not available at the time of the experiment.

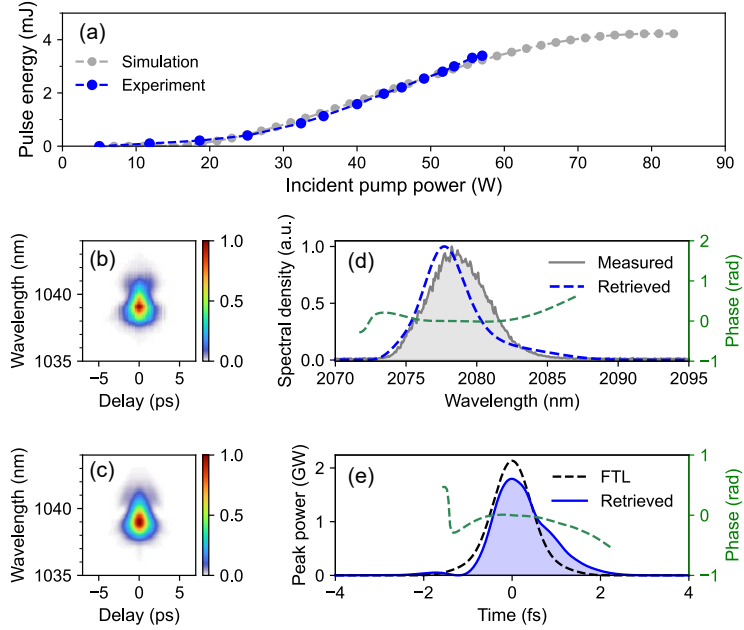

Fig. 6. Output characteristics of Ho:CALGO RA at 1 kHz (a) simulated (grey) and measured (blue) pulse energies as a function of incident pump power, (b) measured, and (c) retrieved FROG traces on a grid of 512×512 and an error of 0.25%. (d) Measured (grey) and retrieved (blue) spectra and the phase (green), and (e) retrieved temporal profile (blue), calculated FTL pulse (black), and the phase(green).

### 3.2 Limitations and strategies for energy scaling and broadband operation

For further power and energy scaling, optimization of doping concentration will be a straightforward approach. It is well known that there are several energy transfer upconversion (ETU) channels in $Ho^{3+}$ ions due to their complex energy levels, which quench upper laser level ions. Since the probability of ETU is proportional to the average distance between ions in the upper level, amplifiers that operate at high inversion are strongly affected by ETU, resulting in reduced conversion efficiency and extra heat generation. Considering conventional high-power Ho:YLF amplifier systems, typically 0.5% doping concentration, i.e., ion density of $6.98\times10^{19}$ $cm^3$ is utilized [12,31], while the ion density of our Ho:CALGO was 1.8 times higher. Therefore, employing a lower doping concentration would improve the amplifier performance, and spectroscopic investigation of the probability of ETU from the $^5I_7$ level and implementation in the simulation model will help optimization. Moreover, other geometries of the gain medium which have superior cooling efficiency, such as thin disk or slab, will overcome the limitation due to thermal load.

Regarding the pulse duration and bandwidth of amplified pulses, they are limited by gain narrowing under our gain factor. Further increasing the seed fluence and reducing the gain factor would be the most promising way to achieve sub-ps pulse duration with mJ-level pulse energy. Not only would it mitigate gain narrowing, but it would also increase the bifurcation threshold, allowing to obtain stable multi-mJ pulses with sub-ps duration. The seed energy is currently limited by available pump diode power in the seed laser oscillator. This can be improved by installing a higher-power pump. Moreover, further optimizing the oscillator emission wavelength and output energy can be achieved, for example, by considering better cooling geometries or other gain materials.

## 4. Conclusion

In this study, we explored the energy scaling of a Ho:CALGO RA system at various repetition rates and operation regimes. An upgraded seed source improved the amplification efficiency due to the higher seed fluence within the amplifier gain bandwidth. The numerical simulation allowed us to predict bifurcation instability and to find the operation regimes for efficient energy extraction at different repetition rates. In the experiment, 10-W average power with sub-ps pulse duration was obtained at 30, 40, and 50-kHz repetition rates, corresponding to pulse energies of 333, 250, and 200 μJ, respectively. Finally, we achieved 3.4-mJ pulse amplification at 1 kHz, reaching a peak power of 1.77 GW with a pulse duration of 1.23 ps. The tunable repetition rate and pulse energy of the developed system enable covering a wide range of applications. For example, higher repetition rates are useful for fast acquisition in many measurements, and energetic pulses can be used for applications that require high intensity. Moreover, the shorter pulse duration of this Ho:CALGO-based system has an advantage for post-compression techniques. In the future, we can potentially directly reach sub-100-fs range with a simple, single compression stage, keeping the pulse spatial and temporal pulse quality and enhancing peak power efficiently. This work paves the way for the realization of a scalable broadband source and applications in the 2.1-μm wavelength range.

**Funding.** DFG Open Access Publication Funds of the Ruhr-Universität Bochum; Deutsche Forschungsge324 meinschaft (EXC-2033, 390677874); European Research Council (ERC) (805202, 101138967) Deutsche Forschungsgemeinschaft (DFG, German Research Foundation) (390677874)

**Acknowledgment.** We acknowledge support by the DFG Open Access Publication Funds of the Ruhr-Universität Bochum. Funded by the Deutsche Forschungsgemeinschaft (DFG, German Research Foundation) under Germanys Excellence Strategy – EXC-2033 – Projektnummer 390677874 - RESOLV. These results are part of a project that has received funding from the European Research Council (ERC) under the European Union's Horizon 2020 research and innovation programme (grant agreement No. 805202 - Project Teraqua). These results are part of a project that has received funding from the European Research Council (ERC) under the European Union's HORIZON-ERC-POC programme (Project 101138967 - Giga2u).

**Disclosures.** The authors declare no conflicts of interest.

**Data availability.** Data underlying the results presented in this paper are available in Ref. [32].

### References

1. I. Astrauskas, B. Považay, A. Baltuška, *et al.*, "Influence of 2.09-μm pulse duration on through-silicon laser ablation of thin metal coatings," Opt. Laser Technol. **133**, 106535 (2021).
2. R. Malik, B. Mills, J. H. V. Price, *et al.*, "Determination of the mid-IR femtosecond surface-damage threshold of germanium," Appl. Phys. A **113**, 127–133 (2013).
3. K.-H. Hong, C.-J. Lai, J. P. Siqueira, *et al.*, "Multi-mJ, kHz, 2.1 μm optical parametric chirped-pulse amplifier and high-flux soft x-ray high-harmonic generation," Opt. Lett. **39**, 3145 (2014).
4. M.-C. Chen, C. Mancuso, C. Hernández-García, *et al.*, "Generation of bright isolated attosecond soft X-ray pulses driven by multicycle midinfrared lasers," Proc. Natl. Acad. Sci. U.S.A. **111**, (2014).
5. C. Gollner, M. Shalaby, C. Brodeur, *et al.*, "Highly efficient THz generation by optical rectification of mid-IR pulses in DAST," APL Photonics **6**, 046105 (2021).
6. M. F. Seeger, D. Kammerer, J. Blöchl, *et al.*, "49 W carrier-envelope-phase-stable few-cycle 2.1 μm OPCPA at 10 kHz," Opt. Express **31**, 24821 (2023).
7. T. Feng, A. Heilmann, M. Bock, *et al.*, "27 W 2.1 μm OPCPA system for coherent soft X-ray generation operating at 10 kHz," Opt. Express **28**, 8724 (2020).
8. C. Gaida, M. Gebhardt, T. Heuermann, *et al.*, "Ultrafast thulium fiber laser system emitting more than 1 kW of average power," Opt. Lett. **43**, 5853 (2018).
9. J. Tang, X. Hua, M. Wu, *et al.*, "111 W picosecond Ho:YAG amplifier and its application to 10 μm mid-infrared generation," Opt. Lett. **50**, 6161 (2025).
10. B. M. Walsh, N. P. Barnes, M. Petros, *et al.*, "Spectroscopy and modeling of solid state lanthanide lasers: Application to trivalent $Tm^{3+}$ and $Ho^{3+}$ in $YLiF_4$ and $LuLiF_4$," J. Appl. Phys. **95**, 3255–3271 (2004).
11. R. L. Aggarwal, D. J. Ripin, J. R. Ochoa, *et al.*, "Measurement of thermo-optic properties of $Y_3Al_5O_{12}$, $Lu_3Al_5O_{12}$, $YAlO_3$, $LiYF_4$, $LiLuF_4$, $BaY_2F_8$, $KGd(WO_4)_2$, and $KY(WO_4)_2$ laser crystals in the 80–300K temperature range," J. Appl. Phys. **98**, 103514 (2005).

12. M. Bock, D. Ueberschaer, M. Mero, *et al.*, "2.05 µm CPA delivering 75 mJ pulses with 2.2 ps duration at a 1 kHz repetition rate," Opt. Express **33**, 17245 (2025).
13. U. Elu, T. Steinle, D. Sánchez, *et al.*, "Table-top high-energy 7 µm OPCPA and 260 mJ Ho:YLF pump laser," Opt. Lett. **44**, 3194 (2019).
14. W. Yao, Y. Wang, S. Tomilov, *et al.*, "8.7-W average power, in-band pumped femtosecond Ho:CALGO laser at 2.1 µm," Opt. Express **30**, 41075 (2022).
15. W. Yao, Y. Wang, S. Ahmed, *et al.*, "Low-noise, 2-W average power, 112-fs Kerr-lens mode-locked Ho:CALGO laser at 2.1 µm," Opt. Lett. **48**, 2801–2804 (2023).
16. J. Zhu, J. Song, Y. Peng, *et al.*, "1 kHz, 10 mJ, sub-200 fs regenerative amplifier utilizing a dual-crystal configuration of Yb:CaGdAlO$_4$ featuring exceptional beam quality," Opt. Express **32**, 34408 (2024).
17. Z. Tu, J. Guo, Z. Gan, *et al.*, "High-power regenerative amplifier based on dual-crystal Yb:CALGO configuration," Opt. Express **33**, 19641 (2025).
18. L. S. Petrov, K. Georgiev, D. Velkov, *et al.*, "Multi-millijoule class, high repetition rate, Yb:CALYO regenerative amplifier with sub-130 fs pulses," Opt. Express **31**, 18765 (2023).
19. P. Loiko, F. Druon, P. Georges, *et al.*, "Thermo-optic characterization of Yb:CaGdAlO$_4$ laser crystal," Opt. Mater. Express **4**, 2241 (2014).
20. A. Suzuki, B. Kassai, Y. Wang, *et al.*, "High-peak-power 2.1 µm femtosecond holmium amplifier at 100 kHz," Optica **12**, 534 (2025).
21. M. Grishin, V. Gulbinas, and A. Michailovas, "Dynamics of high repetition rate regenerative amplifiers," Opt. Express **15**, 9434 (2007).
22. P. Loiko, K. Eremeev, C. Liebald, *et al.*, "Polarized Spectroscopy of Ho:CALGO for Ultrafast Lasers," in *High-Brightness Sources and Light-Driven Interactions Congress* (Optica Publishing Group, 2024), p. MTh4C.3.
23. S.-Y. Chen, Y.-F. Li, G. Wang, *et al.*, "A review of 2.1-µm Tm/Ho doped solid-state lasers: From continuous wavelength to nanosecond-pulse emission," Opt. Mater. **151**, 115292 (2024).
24. I. F. Elder and M. J. P. Payne, "Lasing in diode-pumped Tm:YAP, Tm,Ho:YAP and Tm,Ho:YLF," Optics Communications **145**, 329–339 (1998).
25. A. A. Lagatsky, O. L. Antipov, and W. Sibbett, "Broadly tunable femtosecond Tm:Lu$_2$O$_3$ ceramic laser operating around 2070 nm," Opt. Express **20**, 19349 (2012).
26. A. Schmidt, P. Koopmann, G. Huber, *et al.*, "175 fs Tm:Lu$_2$O$_3$ laser at 2.07 µm mode-locked using single-walled carbon nanotubes," Opt. Express **20**, 5313 (2012).
27. J. Dörring, A. Killi, U. Morgner, *et al.*, "Period doubling and deterministic chaos in continuously pumped regenerative amplifiers," Opt. Express **12**, 1759 (2004).
28. P. Kroetz, A. Ruehl, A.-L. Calendron, *et al.*, "Study on laser characteristics of Ho:YLF regenerative amplifiers: Operation regimes, gain dynamics, and highly stable operation points," Appl. Phys. B **123**, 126 (2017).
29. P. Kroetz, A. Ruehl, K. Murari, *et al.*, "Numerical study of spectral shaping in high energy Ho:YLF amplifiers," Opt. Express **24**, 9905 (2016).
30. K. Hasse, T. Calmano, B. Deppe, *et al.*, "Efficient Yb$^{3+}$:CaGdAlO$_4$ bulk and femtosecond-laser-written waveguide lasers," Opt. Lett. **40**, 3552 (2015).
31. M. Hemmer, D. Sánchez, M. Jelínek, *et al.*, "2-µm wavelength, high-energy Ho:YLF chirped-pulse amplifier for mid-infrared OPCPA," Opt. Lett. **40**, 451–454 (2015).
32. A. Suzuki, M. Müller, S. Tomilov, *et al.*, "Scaling of broadband Ho:CALGO regenerative amplifier to multi-mJ pulse energy," Zenodo, 2026, https://doi.org/10.5281/zenodo.18350216